\documentclass[3p,preprint]{elsarticle}

\usepackage{amssymb}
\usepackage{amsmath} % to use aligned environment

\usepackage{amssymb,amsthm,MnSymbol}
\usepackage{color}

\usepackage{algorithm}
\usepackage{algorithmic}

\usepackage{subfig}

\usepackage{amscd}
\begin{document}

\begin{frontmatter}



% \author{Igor Shevchenko\corref{cor1}}
% \cortext[cor1]{Corresponding author at:}
% \ead{i.shevchenko@imperial.ac.uk}
% \address{Department of Mathematics, Imperial College London, Huxley Building, 180 Queen's Gate, London, SW7 2AZ, UK}

% \author[addr1]{}
\author[addr1]{I. Shevchenko\corref{cor1}}
\ead{i.shevchenko@imperial.ac.uk}
\author[addr1]{P. Berloff}
% \author[addr1,addr2]{P. Berloff}
 
%\fntext[label2]{}
\cortext[cor1]{Corresponding author at:}
%\address{Address\fnref{label3}}
%% \fntext[labecl3]{}

% \address{Department of Mathematics, Imperial College London, Huxley Building, 180 Queen's Gate, London, SW7 2AZ, UK.}
\address[addr1]{Department of Mathematics, Imperial College London, Huxley Building, 180 Queen's Gate, London, SW7 2AZ, UK}
% \address[addr2]{Institute of Numerical Mathematics of the Russian Academy of Sciences, Moscow, Russia}

% \dochead{Short communication}

\title{On a probabilistic evolutionary approach to ocean modelling:\\ From Lorenz-63 to idealized ocean models}
% \title{On a probabilistic evolutionary approach to ocean modelling:\quad From Lorenz-63 to general circulation models}

\begin{abstract}
% \linenumbers
% \onehalfspacing % ???
In this study we develop an alternative way to model the ocean reflecting the chaotic nature of ocean flows and uncertainty of ocean models
-- instead of making use of classical deterministic or stochastic differential equations we offer 
a probabilistic evolutionary approach (PEA) that capitalizes on the use of probabilistic dynamics in phase space. 
The main feature of the data-driven version of PEA proposed in this work is that it does not require to know the physics behind the flow dynamics to model it. 
Within the PEA framework we develop two probabilistic evolutionary methods, %under the umbrella term``Probabilistic advection of the image point''.
which are based on probabilistic evolutionary models using quasi time-invariant structures in phase space.

The methods have been tested on complete and incomplete reference data sets generated by the Lorenz 63 system and  
by an idealized multi-layer quasi-geostrophic model. 
% The methpdshave been tested on two reference flows generated by an idealized multi-layer quasi-geostrophic model set up for a horizontally periodic flat-bottom channel
% and by a comprehensive general circulation model of the North Atlantic, respectively. 
The results show that both methods reproduce large- and small-scale features of the reference flow by keeping the probabilistic dynamics 
within the reference phase space (the phase space of the reference flow).
The proposed approach offers appealing benefits and a great flexibility to ocean modellers working with mathematical models and measurements. 
The most remarkable one is that it provides an alternative to the mainstream ocean parameterizations,
requires no modification of existing ocean models, and easy to implement.
Moreover, it does not depend on the nature of input data, and therefore 
can work with both numerically-computed flows and real measurements from different sources (drifters, weather stations, etc.).
\end{abstract}


\end{frontmatter}

% \linenumbers

% \clearpage
% \tableofcontents


\section{Introduction}
The modern ocean modelling utilizes a wide spectrum of tools ranging from observations to using comprehensive ocean models. Most of the latter 
are based on deterministic or stochastic differential equations, and use both the physics- and data-driven paradigms. The majority 
operate in physical space (e.g.,~\cite{Marshall_etal_1997,hycom2007,Danilov_etal_2017,NEMO2022}), 
while some, umbrellaed under the recently proposed hyper-parameterization approach (e.g.,~\cite{SB2021_J1,SB2022_J1,SB2022_J2,SB2022_J3}),  
take advantage of working in phase space. 
In this study we develop an alternative way to model the ocean reflecting the chaotic nature of ocean flows and uncertainty of ocean models
-- instead of making use of classical deterministic or stochastic differential equations we offer 
a probabilistic evolutionary approach (PEA) that capitalizes on the use of probabilistic dynamics in phase space. 
In this study we develop a data-driven version of PEA the main feature of which is that it does 
not require to know the physics behind the flow dynamics to model it. % (in oppose to the classical physics-driven paradigm). 
It is achieved by its data-driven nature and by shifting the focus from the physical to the reference phase space  (the phase space of the reference flow).
The reference flow can be a numerical solution (generated by an ocean model), observational data, or combination of both.
Within the PEA framework we develop two probabilistic evolutionary methods based on probabilistic evolutionary models using quasi time-invariant 
structures in the reference phase space.
% The first method builds upon the joint probability distribution of directional angles and lengths
% of the vectors in the reference phase space, while the second one utilizes the joint probability distribution of the coordinates of the reference vectors themselves.

% The methods have been tested on two reference data sets generated by the Lorenz 63 system and  
% an idealized multi-layer quasi-geostrophic model set up for a horizontally periodic flat-bottom channel, respectively. 
% % We have tested the methods on two reference flows generated by an idealized multi-layer quasi-geostrophic model set up for a horizontally periodic flat-bottom channel
% % and by a comprehensive general circulation model of the North Atlantic, respectively. 
% The results show that the proposed methods reproduce the reference flow 
% dynamics by utilizing quasi time-invariant structures and therefore keeping the probabilistic dynamics within the reference phase space.
%
The PEA offers appealing benefits and a great flexibility to ocean modellers working with mathematical models and measurements:
(1) it requires no modification of existing ocean models, (2) easy to implement, 
and (3) does not depend on the nature of input data, i.e. it can take not only numerically-computed flows as input data
but also real measurements from different sources (drifters, weather stations, etc.), or combination of both.
Most remarkably, the PEA provides an alternative to the mainstream ocean parameterizations. Namely, instead of running long high-resolution simulations of ocean models
one can generate its relatively short coarse-grained version (that retains the nominally-resolved, resolved on the coarse grid, flow features) and use it as input data for the PEA,
thus replacing computationally-intensive ocean models with a way faster probabilistic evolutionary models.
In other words, the PEA addresses the eddy-parameterization problem from a different angle:
it shifts the focus from the physical space to the phase space of the model and consider the inability of the low-resolution model 
to reproduce the nominally-resolved flow structures as the persistent tendency of the phase 
space trajectory representing the low-resolution solution to escape the reference phase space. 
% (the phase space occupied by the reference eddy-resolving solution projected onto the coarse grid). 

% ????????????????????????????????????????????????????????????????????????
% Parameterizations of small (unresolved and under-resolved) processes and their effect on large (resolved) scales 
% can significantly compromise the fidelity of low-resolution ocean 
% models in both idealized and realistic settings. In the physical space of the model, it manifests itself as fields (velocity, temperature, etc.)
% with notably degraded or no-at-all large-scale flows (compared to their reference flows -- high-resolution flows projected onto the coarse grid),
% not to mention the smaller scales that are however resolved on the computational grid.
% In the phase space of the model, the failure of a low-resolution model corresponds to the trajectory escaping the reference phase space 
% (the phase space occupied by the reference solution -- the high-resolution solution projected onto the coarse grid). 
% The mainstream approach to parameterisations (mathematically simple and physically justified approximations of the key unresolved and under-resolved processes) 
% is to parameterize these processes in the physical space
% (e.g.,~\citet{GentMcwilliams1990,DuanNadiga2007,Frederiksen_et_al2012,
% PortaMana_Zanna2014,CooperZanna2015,
% Grooms_et_al2015,Berloff_2015,Berloff_2016,Berloff_2018,Ryzhov_etal_2019,
% CCHWS2019_1,Ryzhov_etal_2020,
% CCHWS2019_3,CCHWS2020_4,CCHPS2020_J2}). 
% ???????????????????????????????????????????????????????????????????????

First, we explain the probabilistic evolutionary approach and how to build 
probabilistic evolutionary models, and show how they work on the example of the Lorenz 63 system with complete 
and incomplete data sets. 
Then, we apply them to multi-layer quasi-geostrophic (QG) flows with complete and incomplete reference data. 
It is worth mentioning that this work is intended as a proof of concept, therefore we deliberately reduce the technicalities beyond the PEA to a bare minimum, while 
focusing the attention on the key points of the approach.

\section{The probabilistic evolutionary approach (PEA)\label{sec:lorenz63}}
\begin{figure}[H]
\centering
\includegraphics[scale=0.22]{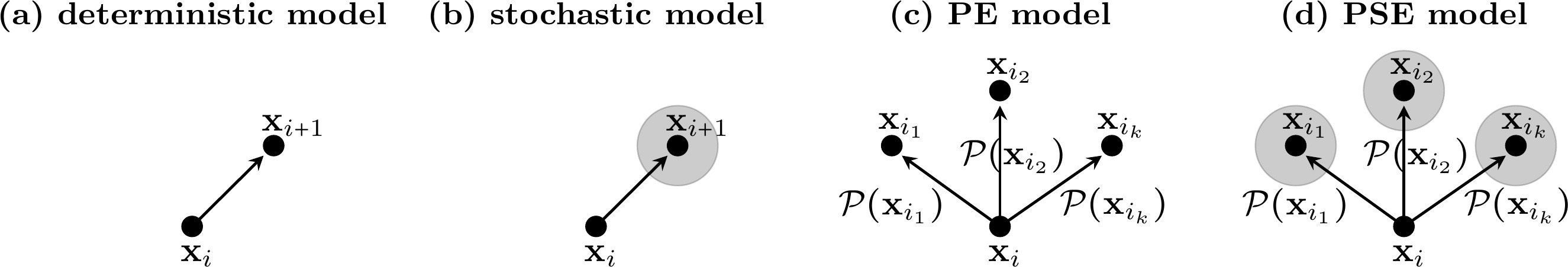}
\caption{Shown is a change of state in {\bf (a)} deterministic models, {\bf (b)} stochastic models, {\bf (c)} probabilistic evolutionary (PE) models, {\bf (d)}
probabilistic-stochastic evolutionary (PSE) models; state and their neighbourhoods are denoted by black dots and grey discs, function $\mathcal{P}(\mathbf{x}_{i_k})$
is a transition probability function for state $\mathbf{x}_{i_k}$.
}
\label{fig:dynamics}
\end{figure}

The probabilistic evolutionary approach is a new approach to ocean modelling that capitalizes on the chaotic nature of ocean dynamics by taking advantage of using
the probability distribution of states in the reference phase space as opposed to making use of deterministic 
% (strictly defined by a set of predetermined past states) 
or stochastic 
%(exploiting additive, multitiplicative, or any other type of noise) 
differential equations. 
By construction, fluid dynamics models can be divided into two classes: deterministic and stochastic. In deterministic models, the dynamics is determined 
by a set of deterministic differential equations, i.e. the transition from a state $\mathbf{x}_i$ to a state $\mathbf{x}_{i+1}$ is unique (Figure~\ref{fig:dynamics}a).
Stochastic models expands the class of deterministic models by introducing noise that does not lead to the unique state $\mathbf{x}_{i+1}$ but 
to a cloud of possible states neighbouring state $\mathbf{x}_{i+1}$. In other words,
stochastic models describe the transit from a state $\mathbf{x}_i$ to a neighbourhood of state $\mathbf{x}_{i+1}$ (Figure~\ref{fig:dynamics}b), 
where the size of the neighbourhood depends on the noise amplitude and the way it is included in the model (additively, multiplicatively, or otherwise).

The probabilistic evolutionary approach offers a different point of view: infinitely many states ($x_{i_1},x_{i_2},\ldots$) 
can be reached from a state $\mathbf{x}_i$, and the transition to 
a particular state is defined by a transition probability function, $\mathcal{P}$, which assigns a probability to any possible transition (Figure~\ref{fig:dynamics}c).
Note that probabilistic evolutionary models do not describe the evolution of a probability function, the evolution in these models is governed by a probability function.
In the data-driven version of PEA proposed in this study, the transition probability function is calculated from available reference data as the current stage
changes (i.e., on the fly). In the physics-driven PEA, we envisage that the transition probability function can be defined from the physics of the studied phenomenon.
Another class of models that naturally follows from the stochastic and probabilistic ones is the probabilistic-stochastic evolutionary models in which 
every possible state is replaced with a cloud of states neighbouring it (Figure~\ref{fig:dynamics}d). 
This gives rise to probabilistic evolutionary models with noise.

In this study we are focused on the data-driven PEA, where the transition probability function is calculated locally from available reference data for every transition 
from one state to another, i.e. a new state of the probabilistic flow evolution is defined by the likelihood of reference states neighbouring to the current state
of the probabilistic evolutionary model.
Within the PEA framework, the probabilistic nature of the flow evolution implies that even very unlikely (rare) events are expected to occur once in a while thus echoing
our observations of extreme weather and climate events. 
More importantly, it allows the probabilistic trajectory %of the image point (also called the representative point that shows the current state of the flow evolution)
to cover regions of the reference phase space that are not presented in the reference data set, but can potentially happen. 

It would be helpful to remind that a system of ordinary differential equations 
\begin{equation}
\mathbf{x}'(t)=\mathbf{F}(\mathbf{x}),\quad \mathbf{x}\in\mathbb{R}^n
\label{eq:ode_x} 
\end{equation}
can be geometrically interpreted as a vector field in the phase space of equation~\eqref{eq:ode_x};
here, the prime denotes a time derivative.
The direction of the vector field at a given point $\mathbf{x}$ is determined %by $n-1$ angles of the vector $\mathbf{F}(\mathbf{x})$
by the vector $\mathbf{F}(\mathbf{x})$
for $\forall \mathbf{x}\in\mathbb{R}^n$. %; these angles are implicitly computed when numerically integrating equation~\eqref{eq:ode_x}. 
Once $\mathbf{F}(\mathbf{x})$ is known, it can be used to calculate a new position of point $\mathbf{x}$ in the phase space.
This idea is used in the hyper-parameterization method ``Advection of the image point''~\citep{SB2021_J1,SB2022_J3}.

\begin{figure}[H]
\centering
\includegraphics[scale=0.22]{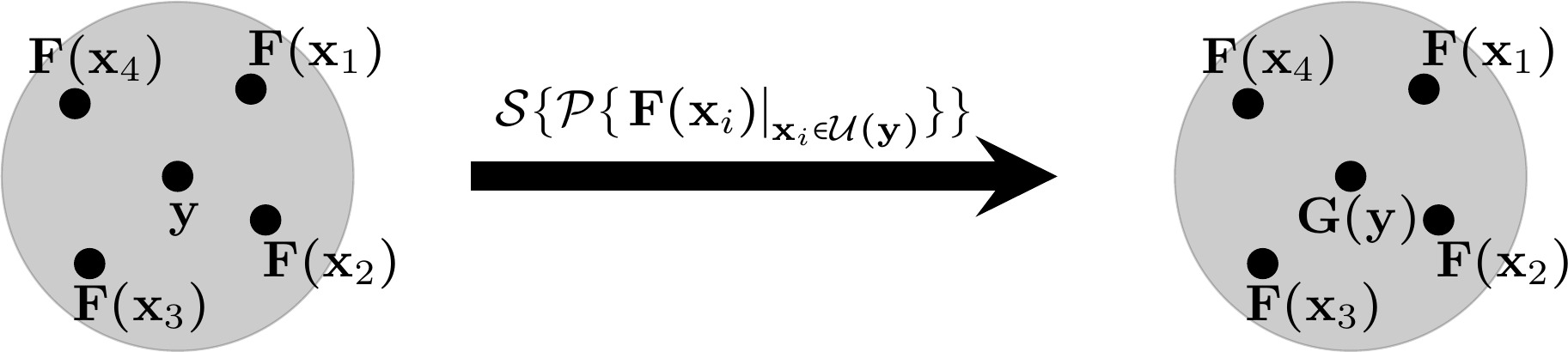}
\caption{Schematic of the data-driven PEA. A state $\mathbf{y}$ of the probabilistic evolutionary model~\eqref{eq:ode_y0} neighbouring 
to the reference vectors $\mathbf{F}(\mathbf{x}_i),\, i=1,2,3,4$. The neighborhood $\mathcal{U}(\mathbf{y})$ of 
$\mathbf{y}$ is denoted by the grey disk, and $\mathcal{S}$ is an operator sampling from $\mathcal{P}$.
}
\label{fig:pea}
\end{figure}

The PEA works in a different way. In the data-driven version of PEA the analytical form of equation~\eqref{eq:ode_x} is not available. 
The only reference data available to the PEA is a numerical solution of~\eqref{eq:ode_x}, i.e. the reference solution, 
or observations if one works with data from weather stations, satellites, etc.
Therefore, instead of directly calculating $\mathbf{F}(\mathbf{x})$ at a give point $\mathbf{x}$, 
the PEA computes a new vector, $\mathbf{G}(\mathbf{y})$, by sampling from the transition probability function, $\mathcal{P}$, 
based on the joint probability distribution of states neighbouring to the current state in the reference phase space (Figure~\ref{fig:pea}).

Thus, the probabilistic evolutionary model can be written as follows:
\begin{equation}
\mathbf{y}'(t)=\mathcal{S}\{\mathcal{P}\{\left.\mathbf{F}(\mathbf{x}(t))\right|_{\mathbf{x}(t)\in\mathcal{U}(\mathbf{y}(t))}\}\},\quad \mathbf{y}(t_0)=\mathbf{x}(t_0),
\label{eq:ode_y0}
\end{equation}
with $\mathcal{S}$ being an operator sampling from the joint probability distribution, 
i.e. it returns a point in phase space given a probability.

We have developed two methods within the PEA framework. These methods use different probability distributions to calculate the 
directional vector $\mathbf{G}(\mathbf{y})$ for the probabilistic evolutionary model.
The first method calculates directional angles and the length of $\mathbf{G}(\mathbf{y})$ 
based on its joint probability distribution function computed from available reference data. 
Hence the probabilistic evolutionary model for the probabilistic solution $\mathbf{y}(t)$ is given by
\begin{equation}
\mathbf{y}'(t)=\mathbf{G}(\mathbf{y})+\mathcal{N}(\mathbf{x}(t),\mathbf{y}(t)),\quad
\mathbf{G}(\mathbf{y}):=C^{-1}_S\{\mathcal{S}\{\mathcal{P}_a\{C_S\{\left.\mathbf{F}(\mathbf{x}(t))\right|_{\mathbf{x}(t)\in\mathcal{U}(\mathbf{y}(t))}\}\}\}\},\quad
\mathbf{y}(t_0)=\mathbf{x}(t_0),
\label{eq:ode_y1}
\end{equation}
where $\mathcal{U}(\mathbf{y}(t))$ is the neighbourhood of probabilistic solution $\mathbf{y}(t)$, 
$C_S$ is the transformation from Cartesian to spherical coordinates (used to compute the angles and lengths of reference vectors), 
$\mathcal{P}_a$ is a transition probability function based on the joint probability distribution of 
directional angles and lengths of the reference vectors 
neighbouring to the current state $\mathbf{y}(t)$ in the phase space of~\eqref{eq:ode_x}, i.e. the vectors 
$\left.\mathbf{F}(\mathbf{x}(t))\right|_{\mathbf{x}(t)\in\mathcal{U}(\mathbf{y}(t))}$; $C^{-1}_S$ is the inverse of $C_S$ used to 
compute $\mathbf{G}(\mathbf{y})$ in the Cartesian space.
The second term on the right hand side of equation~\eqref{eq:ode_y1} is a nudging term:
\begin{equation}
\mathcal{N}(\mathbf{x}(t),\mathbf{y}(t)):=\eta\left(\frac{1}{M}\sum\limits_{i\in\mathcal{U}(\mathbf{y}(t))}\mathbf{x}(t_i)-\mathbf{y}(t)\right),
\label{eq:prob_nudging0}
\end{equation}
where $\eta$ is a nudging strength, $M$ is the number of nearest (in $l_2$ norm) to the solution $\mathbf{y}(t)$ 
points over which the averaged reference solution $\mathbf{x}(t)$ is computed.

The second method does not use the joint probability distribution of the directional angles and lengths of 
reference vectors to compute $\mathbf{G}(\mathbf{y})$. Instead, it computes $\mathbf{G}(\mathbf{y})$
from the joint probability distribution of the coordinates of reference vectors neighbouring 
to the current state $\mathbf{y}(t)$ in the phase space of~\eqref{eq:ode_x}.
Hence, the probabilistic evolutionary equation reads as follows:
\begin{equation}
\mathbf{y}'(t)=\mathbf{G}(\mathbf{y})+\mathcal{N}(\mathbf{x}(t),\mathbf{y}(t)),\quad 
\mathbf{G}(\mathbf{y}):=\mathcal{S}\{\mathcal{P}_c\{\left.\mathbf{F}(\mathbf{x}(t))\right|_{\mathbf{x}(t)\in\mathcal{U}(\mathbf{y}(t))}\}\},\quad
\mathbf{y}(t_0)=\mathbf{x}(t_0),
\label{eq:ode_y2}
\end{equation}
where $\mathcal{P}_c$ is a transition probability function based on the joint probability distribution of 
the coordinates of the reference vectors neighbouring to the current state $\mathbf{y}(t)$ in the phase space of~\eqref{eq:ode_x}, 
i.e. the vectors $\left.\mathbf{F}(\mathbf{x}(t))\right|_{\mathbf{x}(t)\in\mathcal{U}(\mathbf{y}(t))}$.
To sample from $\mathcal{P}_a$ and $\mathcal{P}_c$, we use the sampling operator $\mathcal{S}$ based on 
the inverse transform sampling method~\citep{Devroye1986}. We also tried the rejection sampling
but did not observe that much of a difference.

{\bf Probabilistic nudging}. 
The form of the nudging term used in the probabilistic evolutionary models~\eqref{eq:ode_y1} and~\eqref{eq:ode_y2} 
is governed by our desire to keep the probabilistic evolutionary model as simple as possible.
Different metrics and forms of the nudging term can be used instead (for example, adaptive~\citep{SB2022_J1} or probabilistic nudging).

The idea behind probabilistic nudging is also based on using the probabilistic evolutionary machinery. However,
instead of computing the joint probability distribution of the vectors 
$\left.\mathbf{F}(\mathbf{x}(t))\right|_{\mathbf{x}(t)\in\mathcal{U}(\mathbf{y}(t))}$ as above, we compute  
the joint probability distribution of the states $\mathbf{x}(t)$ themselves. Thus, the probabilistic nudging term can be written as
\begin{equation}
\mathcal{N}(\mathbf{x}(t),\mathbf{y}(t)):=\eta\left(\left.\mathcal{P}_c\{\mathbf{x}(t)\}\right|_{\mathbf{x}(t)\in\mathcal{U}(\mathbf{y}(t))}-\mathbf{y}(t)\right),
\label{eq:prob_nudging}
\end{equation}
where $\mathcal{P}_c$ is a probability function based on the joint probability distribution of
the coordinates of the reference states $\mathbf{x}(t)$  neighbouring to the current state $\mathbf{y}(t)$ in the phase space of~\eqref{eq:ode_x}, 
i.e. the states $\left.\mathbf{x}(t)\right|_{\mathbf{x}(t)\in\mathcal{U}(\mathbf{y}(t))}$. We do not study how the probabilistic nudging
performs in this work and leave it for the future research.

{\bf On the optimal choice of parameters}. 
Note that the neighbourhood $\mathcal{U}(\mathbf{y}(t))$ in equations~\eqref{eq:ode_y1} and~\eqref{eq:ode_y2} is computed as $N$ (and $M$ for the nudging term) nearest (in $l_2$ norm) 
to the solution $\mathbf{y}(t)$ points. The neighbourhood can be 
computed differently, and the way it is computed affects the solution. 
The optimal value of $N$ and $M$ (as well as $\eta$) for a given reference solution can be computed by solving the following optimization problem
\begin{equation}
 \min_{N,M,\eta}\, \mathcal{F}(\mathbf{x}(t),\mathbf{y}(t)),\quad t\in[0,T]
 \label{eq:criterium}
\end{equation}
where $\mathcal{F}$ is a problem-specific function, and $T$ is the length of the reference solution $\mathbf{x}(t)$.
For example, $\mathcal{F}$ can be defined as a norm of the difference between the reference and probabilistic solutions.
Our choice of $N$, $M$, and $\eta$ is driven by our measure of goodness 
(to keep the probabilistic solution $\mathbf{y}(t)$ within the reference phase space).
This measure is used because it allows the probabilistic solution to evolve in the neighborhood of the reference phase space, since
the failure to do so results in a wrong flow dynamics typically shown by low-resolution ocean models.
Studying optimal strategies of computing the neighbourhood and its size as well 
as the nudging strength $\eta$ is a topic beyond the scope of the present paper.

% \newpage
% 
% The probabilistic nature of the method allows it to capture the whole spectrum of events ranging from frequent (with high probability to happen) 
% to rare (with low probability to happen) ones.

% --------------------------------------
% The motion of the image point in phase space is determined by the direction filed of a given system of ordinary differential equations.
% 
% The higher the probability of a particular state the more likely it will be selected to as the next state...
% probabilistic evolution, i.e. the current state is advanced in the direction which is more likely
% --------------------------------------

\begin{figure}[H]
\centering
\includegraphics[scale=0.22]{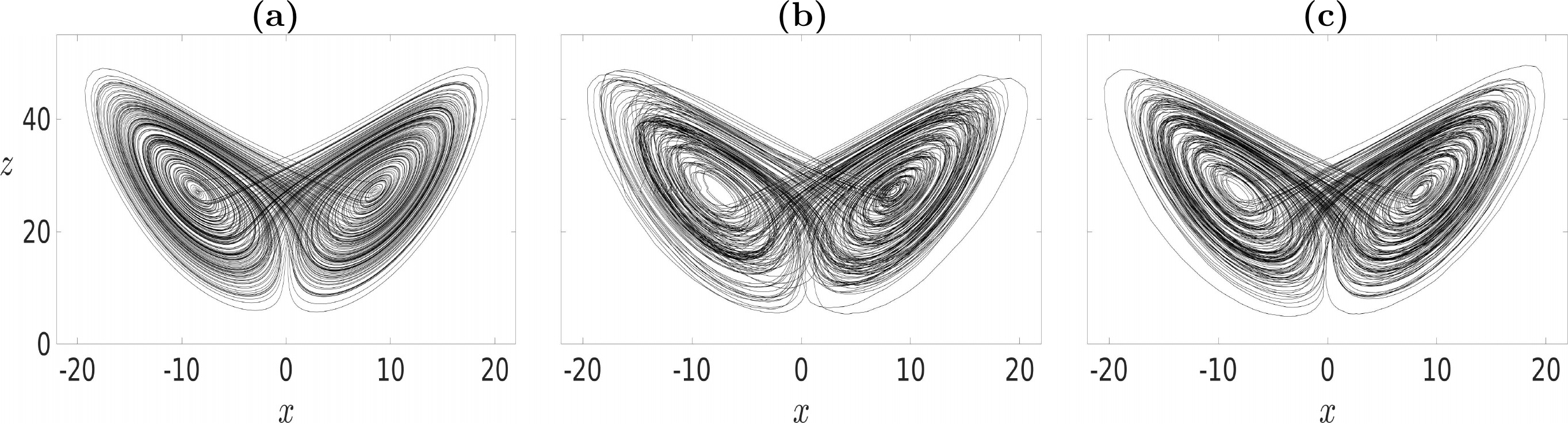}
% \hspace*{-0.75cm}
% \begin{tabular}{ccc}
% \hspace*{1.8cm}\begin{minipage}{0.1\textwidth} {\bf (a)} \end{minipage} & \hspace*{1.8cm}\begin{minipage}{0.1\textwidth} {\bf (b)} \end{minipage}
% & \hspace*{1.8cm}\begin{minipage}{0.1\textwidth} {\bf (c)} \end{minipage}\\
% & & \\[-0.5cm]
% \hspace*{0cm}\begin{minipage}{0.3\textwidth}\vspace*{0.0cm}\includegraphics[height=4cm,width=5.25cm]{{lorenz63_t200}.jpg}\end{minipage} &
% \hspace*{0cm}\begin{minipage}{0.3\textwidth}\vspace*{0.0cm}\includegraphics[height=4cm,width=5.25cm]{{lorenz63_t200_deg2_mm10_eta0_l2squared_prob_angles}.jpg}\end{minipage} &
% \hspace*{0cm}\begin{minipage}{0.3\textwidth}\vspace*{0.0cm}\includegraphics[height=4cm,width=5.25cm]{{lorenz63_t200_deg2_mm10_eta0_l2squared_prob_coords}.jpg}\end{minipage}\\
% \end{tabular}
\caption{Shown is {\bf (a)} the solution of the Lorenz system~\eqref{eq:Lorenz63} for the time interval $t\in[0,200]$, 
{\bf (b)} and {\bf (c)} the probabilistic solutions of~\eqref{eq:ode_y1} and~\eqref{eq:ode_y2}, respectively.
The probabilistic solutions use only the first half of the reference data (i.e., $t\in[0,100]$), 
and over the second half, $t\in[100,200]$, the probabilistic evolutionary methods work out of the sample. 
Both probabilistic solutions stay in the reference phase space, and reproduce the Lorenz attractor.}
\label{fig:lorenz63}
\end{figure}

As an example, we consider the Lorenz 63 system~\citep{Lorenz1963}:
\begin{equation}
\mathbf{x}'(t)=\mathbf{F}(\mathbf{x}(t)),\quad \mathbf{F}:=
\begin{pmatrix}
\sigma(y-x)\\
x(\rho-z)-y\\
xy-\beta z\\
\end{pmatrix},
\label{eq:Lorenz63}
\end{equation}
with $\mathbf{x}(t)=(x(t),y(t),z(t))$, and $\sigma=10$, $\beta=8/3$, $\rho=28$. As an initial condition,
we take $\mathbf{x}(t_0)=(-8.6,-12.4,21.0)$ to make sure 
the solution is close to the Lorenz attractor (Figure~\ref{fig:lorenz63}a).
Along with the solution of the Lorenz system, we compute the probabilistic solutions
to equations~\eqref{eq:ode_y1} and~\eqref{eq:ode_y2} with $N=10$ (Figures~\ref{fig:lorenz63}b,c); we have also tested the probabilistic evolutionary methods for $N=5$ 
and report that the results are qualitatively the same (not shown).
Note that all probabilistic solutions are computed without nudging (i.e., for $\eta=0$), as it is not required to reproduce the Lorenz attractor;
however, nudging will play an essential role in simulations of the QG model discussed in Section~\ref{sec:qg}. 

As seen in Figures~\ref{fig:lorenz63}b,c, the probabilistic solution %for different sizes of the neighbourhood $N$ and
stay within the same region of the phase space 
as the reference solution of the Lorenz system, despite that only the first half of the reference solution is available (i.e., $t\in[0,100]$), 
and the method runs out of the sample over the second half, i.e. for $t\in[100,200]$. This is important, as for the probabilistic evolutionary approach 
the measure of goodness is how close the probabilistic solution is to the reference phase space.

{\bf Incomplete reference data}. 
As the proposed probabilistic evolutionary approach is intended to work with ocean models, it is instructive to 
test its methods on incomplete reference data sets which are not uncommon in ocean modelling, especially when using comprehensive ocean models.
We consider three test cases: (1) gappy dynamics, (2) holey attractor, and (3) disjoint wings.

\begin{figure}[H]
\centering
\includegraphics[scale=0.22]{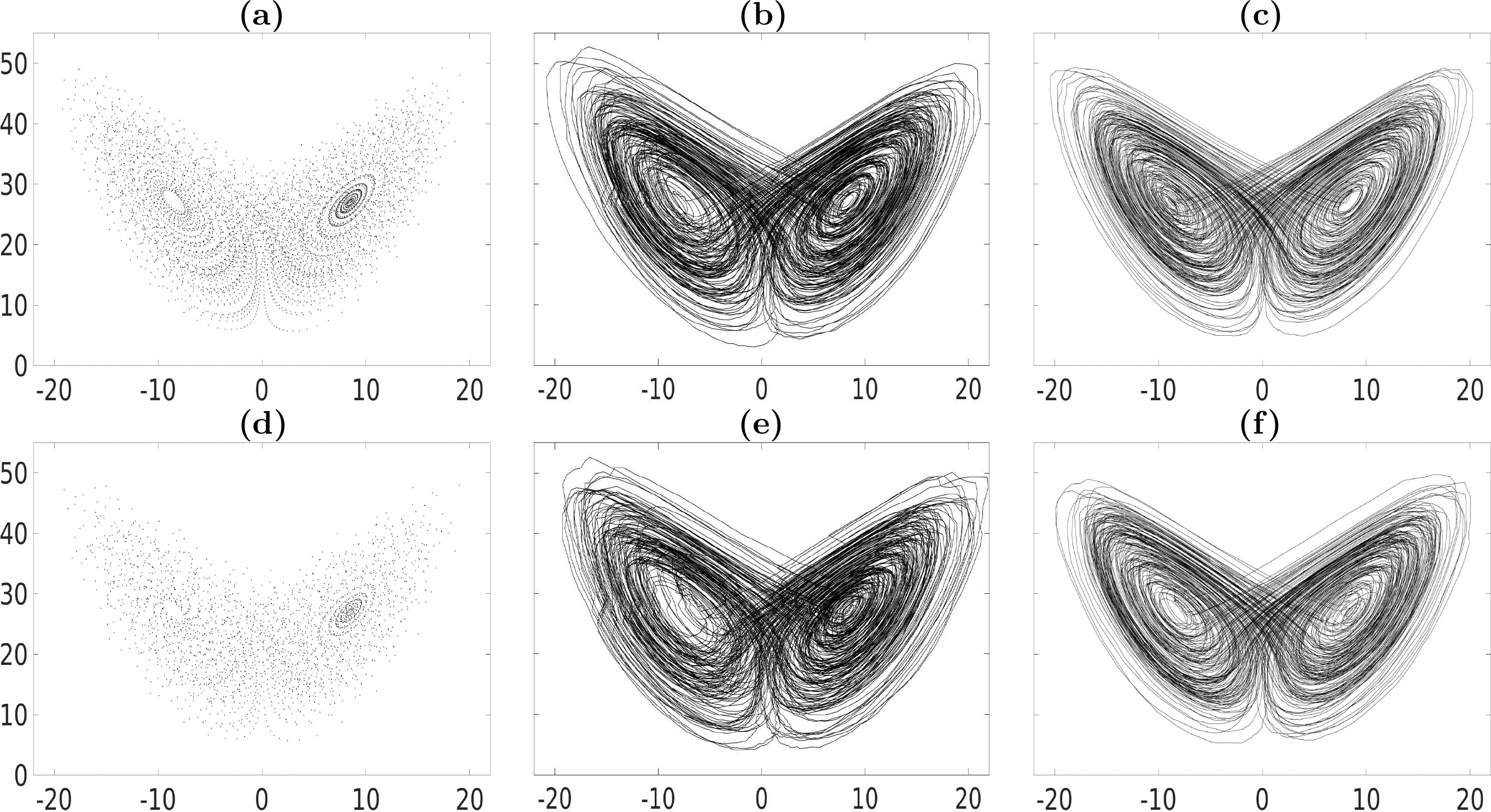}
\caption{Shown is {\bf (a)} the reference solution (over $t\in[0,100]$) with every second point retained,
{\bf (b)}/{\bf (c)} the first/second probabilistic solution over $t\in[0,200]$;
{\bf (d)}-{\bf (f)} are the same as {\bf (a)}-{\bf (c)} but for the reference solution with every forth point retained.
The axes are the same as in Figure~\ref{fig:lorenz63}.}
\label{fig:gappy_dynamics}
\end{figure}

{\bf Gappy dynamics}. For this test we generate two data sets that are used as reference data for both methods.
Namely, we take the Lorenz solution over the time period $[0,100]$ and keep 
every second and every fourth point of the solution, thus retaining only 50\% and 25\% of the original reference data, respectively.
We set $N=10$ and present the results in Figure~\ref{fig:gappy_dynamics}; the results for $N=5$ are qualitatively the same (not shown).
As we see in Figure~\ref{fig:gappy_dynamics}, both methods keep the probabilistic solution on the Lorenz attractor, 
despite the reference solution is missing a significant portion of data.

\begin{figure}[H]
\centering
\includegraphics[scale=0.22]{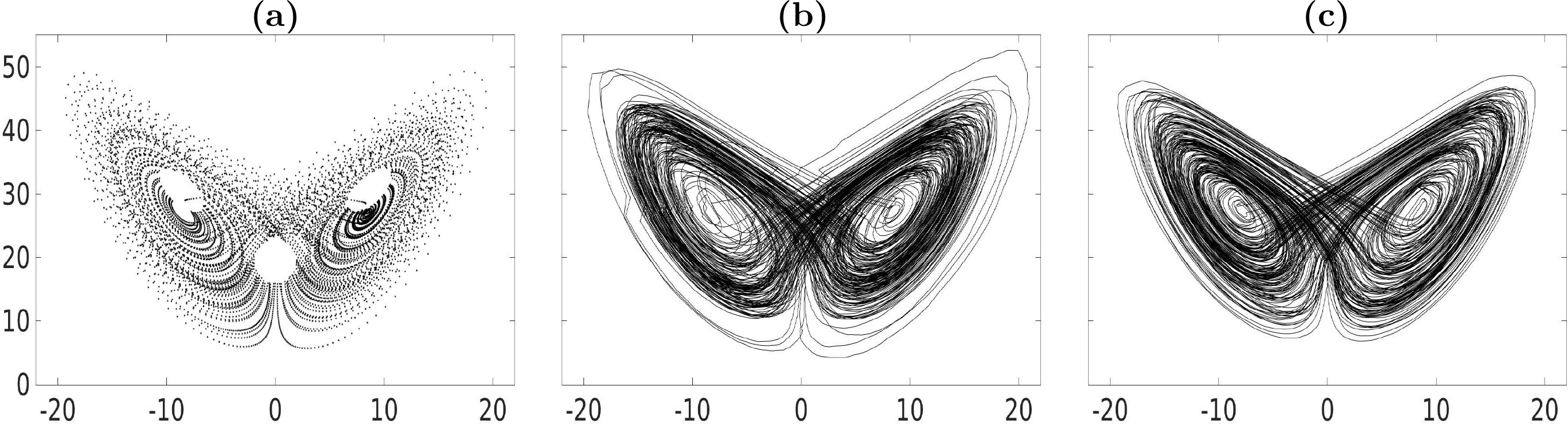}
% \hspace*{-0.75cm}
% \begin{tabular}{ccc}
% \hspace*{1.8cm}\begin{minipage}{0.1\textwidth} {\bf (a)} \end{minipage} & \hspace*{1.8cm}\begin{minipage}{0.1\textwidth} {\bf (b)} \end{minipage}
% & \hspace*{1.8cm}\begin{minipage}{0.1\textwidth} {\bf (c)} \end{minipage}\\
% & & \\[-0.5cm]
% \hspace*{0cm}\begin{minipage}{0.3\textwidth}\vspace*{0.0cm}\includegraphics[height=4cm,width=5.25cm]{{lorenz63_refsol_t100__holey_attractor_r4}.jpg}\end{minipage} &
% \hspace*{0cm}\begin{minipage}{0.3\textwidth}\vspace*{0.0cm}\includegraphics[height=4cm,width=5.25cm]{{lorenz63_t200_deg2_mm10_eta0_l2squared_prob_angles__holey_attractor_r4}.jpg}\end{minipage} &
% \hspace*{0cm}\begin{minipage}{0.3\textwidth}\vspace*{0.0cm}\includegraphics[height=4cm,width=5.25cm]{{lorenz63_t200_deg2_mm10_eta0_l2squared_prob_coords__holey_attractor_r4_h0.0102}.jpg}\end{minipage}\\
% \end{tabular}
\caption{Shown is {\bf(a)} the reference solution (over $t\in[0,100]$) with three holes of radius 4, 
{\bf (b)}/{\bf (c)} the first/second probabilistic solution over $t\in[0,200]$. 
The axes are the same as in Figure~\ref{fig:lorenz63}.}
\label{fig:holey_attractor}
\end{figure}

{\bf Holey attractor}. This test is harder compared to the first one, as we cut out some regions of the reference dynamics by making three holes of radius 4  
in the attractor itself (Figure~\ref{fig:holey_attractor}a). As in the first test, both probabilistic evolutionary methods (Figures~\ref{fig:holey_attractor}b,c) 
keep the solution on the attractor. More importantly, the methods restore the dynamics on the attractor as if there are no holes. 
 
\begin{figure}[H]
\centering
\includegraphics[scale=0.22]{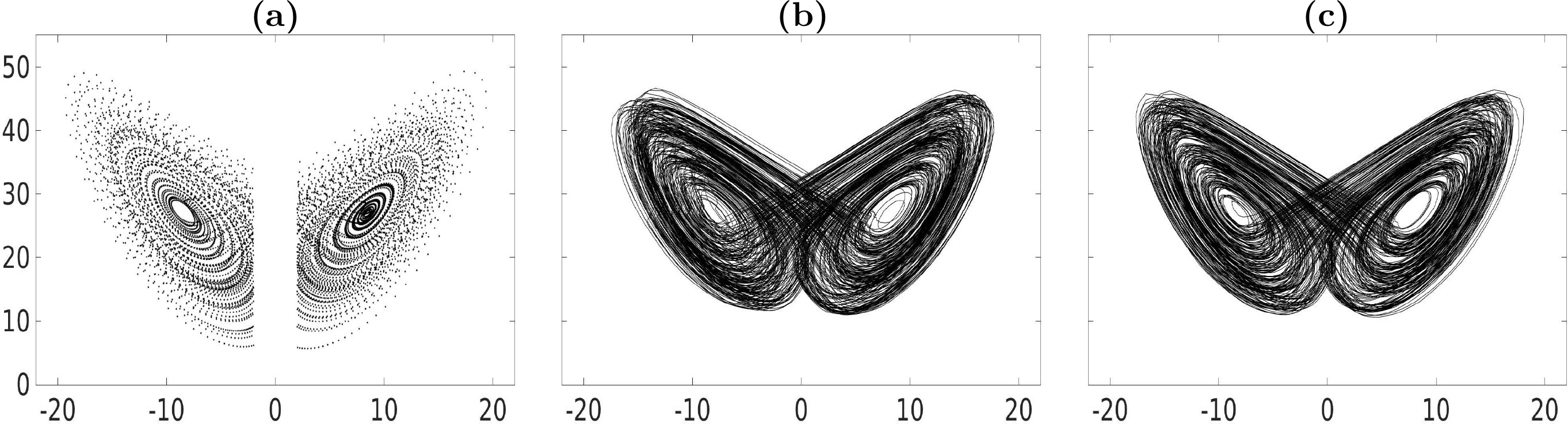}
% \hspace*{-0.75cm}
% \begin{tabular}{ccc}
% \hspace*{1.8cm}\begin{minipage}{0.1\textwidth} {\bf (a)} \end{minipage} & \hspace*{1.8cm}\begin{minipage}{0.1\textwidth} {\bf (b)} \end{minipage}
% & \hspace*{1.8cm}\begin{minipage}{0.1\textwidth} {\bf (c)} \end{minipage}\\
% & & \\[-0.5cm]
% \hspace*{0cm}\begin{minipage}{0.3\textwidth}\vspace*{0.0cm}\includegraphics[height=4cm,width=5.25cm]{{lorenz63_refsol_t100__disjoint_wings_r2}.jpg}\end{minipage} &
% \hspace*{0cm}\begin{minipage}{0.3\textwidth}\vspace*{0.0cm}\includegraphics[height=4cm,width=5.25cm]{{lorenz63_t200_deg2_mm10_eta0_l2squared_prob_angles__disjoint_wings_r2_dtcoeff1.3}.jpg}\end{minipage} &
% \hspace*{0cm}\begin{minipage}{0.3\textwidth}\vspace*{0.0cm}\includegraphics[height=4cm,width=5.25cm]{{lorenz63_t200_deg2_mm10_eta0_l2squared_prob_coords__disjoint_wings_r2_dtcoeff1.3}.jpg}\end{minipage}\\
% \end{tabular}
\caption{Shown is {\bf(a)} the reference solution (over $t\in[0,100]$) with disjoint wings, 
{\bf (b)}/{\bf (c)} the first/second probabilistic solution over $t\in[0,200]$. 
The axes are the same as in Figure~\ref{fig:lorenz63}.}
\label{fig:disjoint_wings}
\end{figure}

{\bf Disjoint wings}. It is the hardest test in the series, as we cut the attractor into two disjoint sets (Figure~\ref{fig:disjoint_wings}a) 
thus simulating a significantly corrupted data set; the cut width is 2.
As seen in Figures~\ref{fig:disjoint_wings}b,c, both probabilistic evolutionary methods not only recover the attractor but also restore the dynamics in between the wings where the reference
solution is unavailable.

\begin{figure}[H]
\centering
\includegraphics[scale=0.22]{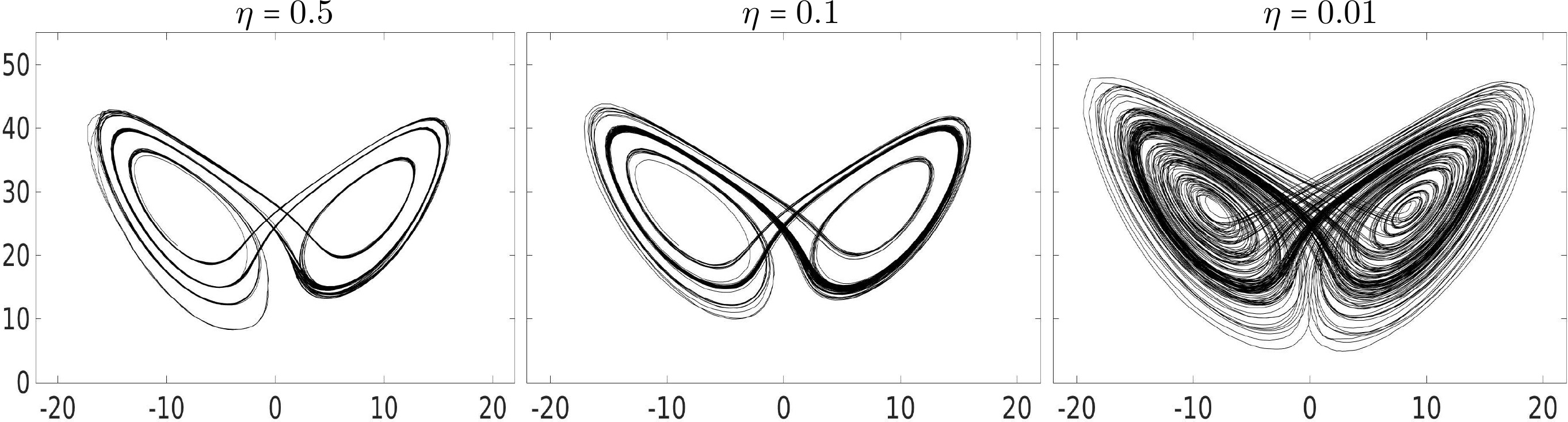}
% \hspace*{-0.75cm}
% \begin{tabular}{ccc}
% \hspace*{1.5cm}\begin{minipage}{0.1\textwidth} $\eta=0.5$ \end{minipage} & \hspace*{1.1cm}\begin{minipage}{0.1\textwidth} $\eta=0.1$ \end{minipage}
% & \hspace*{1.0cm}\begin{minipage}{0.1\textwidth} $\eta=0.01$ \end{minipage}\\
% & & \\[-0.5cm]
% \hspace*{0cm}\begin{minipage}{0.3\textwidth}\vspace*{0.0cm}\includegraphics[height=4cm,width=5.25cm]{{lorenz63_nudging_eta0.5}.jpg}\end{minipage} &
% \hspace*{0cm}\begin{minipage}{0.3\textwidth}\vspace*{0.0cm}\includegraphics[height=4cm,width=5.25cm]{{lorenz63_nudging_eta0.1}.jpg}\end{minipage} &
% \hspace*{0cm}\begin{minipage}{0.3\textwidth}\vspace*{0.0cm}\includegraphics[height=4cm,width=5.25cm]{{lorenz63_nudging_eta0.01}.jpg}\end{minipage}\\
% \end{tabular}
\caption{Shown is the dependence of the Lorenz solution (computed with the second probabilistic evolutionary method) on the nudging strength $\eta$ for $t\in[0,200]$.
The results for the first method are qualitatively similar (not shown).}
\label{fig:lorenz63_nudging}
\end{figure}

{\bf On a detrimental role of nudging}. We did not use nudging in the probabilistic evolutionary methods to model the Lorenz system,
as it is not necessary to reproduce the reference dynamics. But,
within the context of the QG model discussed below we will see the beneficial effect of nudging on the flow dynamics.
However, it should be noted in advance that nudging can also play a detrimental role when using out of place. As an example, we take the Lorenz system and demonstrate 
how improper use of nudging can affect the solution. As seen in Figure~\ref{fig:lorenz63_nudging}, the nudging strength has a significant effect on the solution.
Namely, for a relatively strong nudging the probabilistic solution cannot properly develop on the attractor, and the whole dynamics is confined in a narrow strip.
It is a simple but illustrative example of how one should use caution to properly adjust the nudging strength to get a good solution.

Summing up our findings for the Lorenz 63 system, we conclude that the proposed probabilistic evolutionary methods have strong potential for modelling geophysical flows.
It may seem that the first method is somewhat inferior to the second one, as the former
gives the trajectories which do stay in the reference phase space, but can experience 
rapid changes of the direction that may lead to undesirable effects in geophysical flows;   
in fact, these changes can be mitigated by using a shorter time step (not shown). 
%It happens because the directional angles are more sensitive to the computational error than the coordinates of the directional vector itself.
Despite this seeming disadvantage we does not disregard the first method and test it too on the QG model, as 
this effect might be of minor or no influence within the context of QG dynamics.

\section{Multilayer quasi-geostrophic equations\label{sec:qg}}
In this section we apply the probabilistic evolutionary methods to the 2-layer quasi-geostrophic (QG) model describing the evolution of potential vorticity (PV) anomaly 
$\mathbf{q}=(q_1,q_2)$ in a domain $\Omega$~\citep{Pedlosky1987}:
\begin{equation}
\begin{aligned}
\partial_t q_1+\mathbf{u}_1\cdot\nabla q_1&=\nu\nabla^4\psi_1-\beta\partial_x \psi_1,\\
\partial_t q_2+\mathbf{u}_2\cdot\nabla q_2&=\nu\nabla^4\psi_2-\mu\nabla^2\psi_2-\beta\partial_x \psi_2,
\end{aligned}
\label{eq:pv}
\end{equation}
where $\mathbf{u}=(u,v)$ is a velocity vector, $\boldsymbol{\psi}=(\psi_1,\psi_2)$ is the stream function in the top and bottom layers, 
$\nu=3.125\,\rm m^2 s^{-1}$ is the lateral eddy viscosity, $\beta=2\times10^{-11}\, {\rm m^{-1}\, s^{-1}}$ is the planetary vorticity gradient, and
$\mu=4\times10^{-8}\, {\rm s^{-1}}$ is the bottom friction parameter.
The computational domain $\Omega=[0,L_x]\times[0,L_y]\times[0,H]$ is a horizontally periodic flat-bottom channel of depth
$H=H_1+H_2$ given by two stacked isopycnal fluid layers of depth $H_1=1.0\, \rm km$, $H_2=3.0\, \rm km$, and
$L_x=3840\, \rm km$, $L_y=L_x/2$.

Forcing in~\eqref{eq:pv} is introduced via a vertically sheared, baroclinically unstable background flow (e.g.,~\cite{CCHPS2020_J2}):
\begin{equation}
\psi_i\rightarrow-U_i\,y+\psi_i,\quad i=1,2, 
\label{eq:forcing}
\end{equation}
with the background-flow zonal velocities $U=[6.0,0.0]\,\rm cm\, s^{-1}$. 

The PV anomaly and stream function are related through the system of elliptic equations:
\begin{subequations}
\begin{align}
q_1=\nabla^2\psi_1+s_1(\psi_2-\psi_1),\\
q_2=\nabla^2\psi_2+s_2(\psi_1-\psi_2),
\end{align}
\label{eq:q_psi}
\end{subequations}
with the stratification parameters $s_1=4.22\cdot10^{-3}\,\rm km^{-2}$, $s_2=1.41\cdot10^{-3}\,\rm km^{-2}$; chosen so that the first Rossby deformation radius is $Rd_1=25\, {\rm km}$.

System~(\ref{eq:pv})-(\ref{eq:q_psi}) is augmented by the integral mass conservation constraint~\citep{McWilliams1977}:
\begin{equation}
\partial_t\iint\limits_{\Omega}(\psi_1-\psi_2)\ dydx=0,
\label{eq:masscon}
\end{equation}
by the periodic horizontal boundary conditions set at eastern, $\Gamma_2$, and western, $\Gamma_4$, boundaries
\begin{equation}
\boldsymbol{\psi}\Big|_{\Gamma_2}=\boldsymbol{\psi}\Big|_{\Gamma_4}\,,\quad \boldsymbol{\psi}=(\psi_1,\psi_2)\,,
\label{eq:bc24}
\end{equation}
and no-slip boundary conditions 
\begin{equation}
\boldsymbol{u}\Big|_{\Gamma_1}=\boldsymbol{u}\Big|_{\Gamma_3}=0\,.
\label{eq:bc13}
\end{equation}
set at northern, $\Gamma_1$, and southern, $\Gamma_3$, boundaries of the domain $\Omega$.

The QG equations~\eqref{eq:pv} can be recast in the form of equation~\eqref{eq:ode_x} as follows:
\begin{equation}
\mathbf{q}'(t)=\mathbf{F}(\mathbf{q},\boldsymbol{\psi},\mathbf{u}),
\label{eq:pv1} 
\end{equation}
where the right hand side $\mathbf{F}$ defines the vector field used to evolve $\mathbf{q}$;
$\mathbf{F}$ is computed  with the central finite difference in time from the available reference data.
The only difference with the Lorenz system~\eqref{eq:Lorenz63} is the vector $\mathbf{F}$ and the advected quantity $\mathbf{q}$;
note that the dimension of phase space is defined by the number of degrees of freedom used to 
discretize the equation in space. Thus, the analogue of equations~\eqref{eq:ode_y1} and~\eqref{eq:ode_y2} for the QG equations reads:
\begin{equation}
\mathbf{y}'(t)=\mathbf{G}(\mathbf{y})+\mathcal{N}(\mathbf{q}(t),\mathbf{y}(t)),\quad
\mathbf{y}(t_0)=\mathbf{q}(t_0).
\label{eq:ode_q1}
\end{equation}

For the purpose of this study both high- and low-resolution solutions are needed. We compute these solutions 
on uniform grids of size $513\times257$ and $129\times65$ over a period of 4 years after a 10-year initial spin up. 
In order to test the probabilistic evolutionary methods in different regimes, we use both the high-resolution solution and its 
point-to-point projection onto the coarse grid $129\times65$;
the coarse-graining is of little importance to the probabilistic evolutionary approach, and any other method (e.g., interpolation schemes, spatial averaging, or filters) can be used.
The high-resolution solution is needed to study to what extent the probabilistic evolutionary methods can be used as an alternative to high-resolution ocean simulations, whereas
its low-resolution projection is to compare the performance of the methods with the low-resolution solution computed on the coarse grid 
with the QG model. We should also note that these solutions are denoted as $\mathbf{x}(t)$ in equation~\eqref{eq:ode_x},
while their corresponding probabilistic solutions are denoted as $\mathbf{y}(t)$ in equations~\eqref{eq:ode_y1} and~\eqref{eq:ode_y2}.
It is an important comparison which will show whether the probabilistic evolutionary methods can reproduce flow features that 
are presented in the low-resolution projection but are missing in the low-resolution QG solution.
For the purpose of this study, it is enough to consider the first layer PV anomaly, 
as it is much more energetic than the second layer and full of both large- and small-scale flow features.

\begin{figure}[H]
\centering
\includegraphics[scale=0.22]{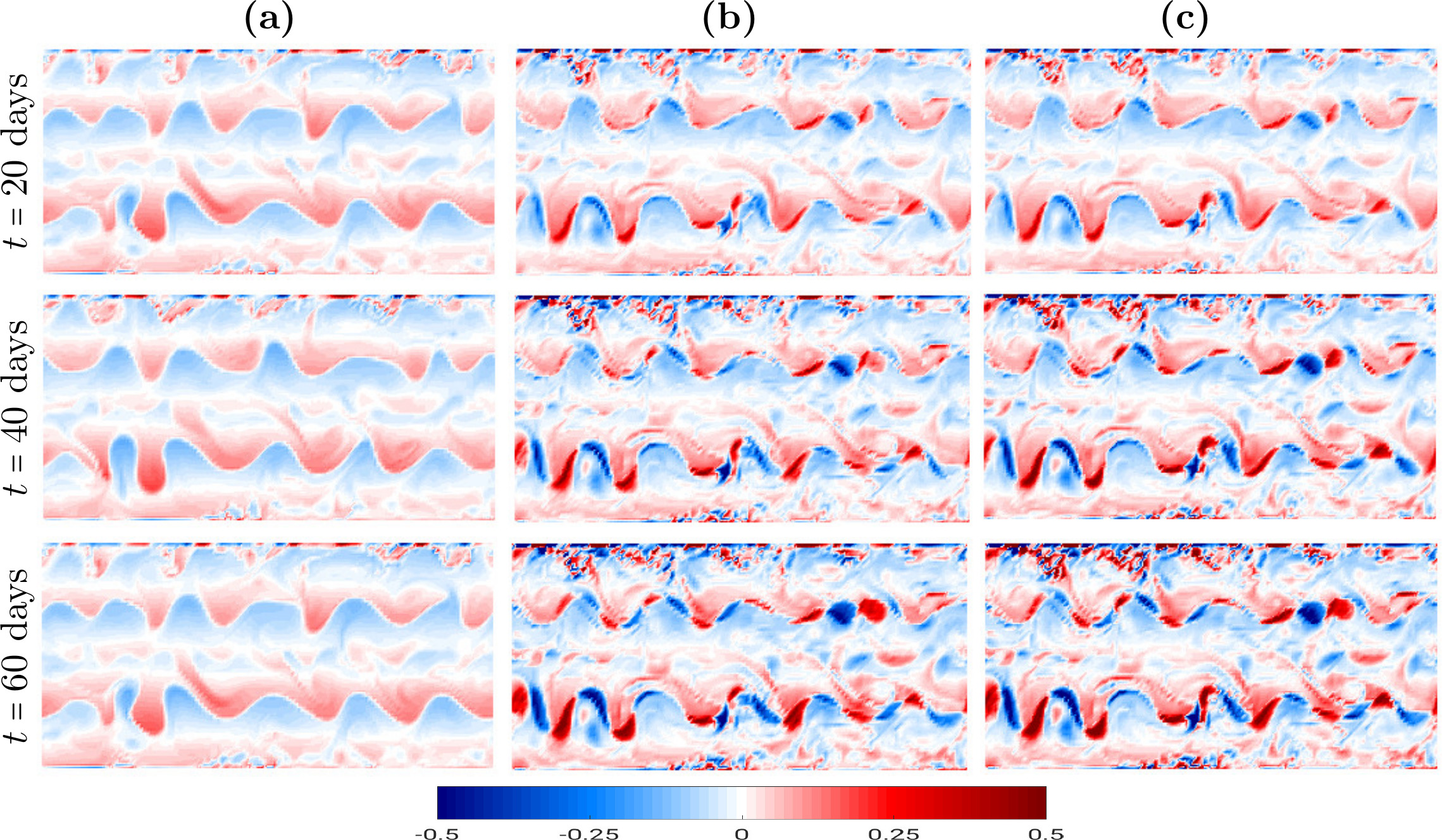}
\caption{Shown are snapshots of the top layer PV anomaly: {\bf (a)} the reference solution (computed on the $513\times257$
grid and then projected onto the $129\times65$ grid), {\bf (b)}/{\bf (c)} probabilistic solution computed with the first/second method. 
All solutions are given in units of $[s^{-1}f^{-1}_0]$, where $f_0=0.83\times10^{-4}\, {\rm s^{-1}}$ is the Coriolis parameter.
Both methods reproduce nominally-resolved on the coarse grid flow features but over a short period of time 
the build-up effect kicks in and arrests the dynamics.}
\label{fig:qg_sol_no_nudging}
\end{figure}

In order to demonstrate the ability of the probabilistic evolutionary methods to reproduce nominally-resolved on the coarse grid flow features,
we take only the first two years of the 4-year long high-resolution solution, coarse-grain it onto the grid of size $129\times65$ and then use it
as a reference solution (Figure~\ref{fig:qg_sol_no_nudging}a).
As for the Lorenz 63, we firstly apply the methods without nudging. 

{\bf The build-up effect}. As seen in Figures~\ref{fig:qg_sol_no_nudging}b,c,
at the very beginning the probabilistic solution reproduces
both large-scale flow structures (two zonally-elongated jets) as well as small-scale vortices and meanders along the jets of the reference solution (Figure~\ref{fig:qg_sol_no_nudging}a).
It is because the probabilistic solution remains in the reference phase space. 
But, after a short period of time the build-up effect takes over and makes the probabilistic solution 
drift away from the reference phase space,
and eventually to settle to a constant in time field (40- and 60-day solutions already show almost no change in time).
It takes only 60 days for the solution to ``freeze'' down to a point of no use. 
% As a result, it leads to a gradual degradation of the probabilistic solution down to a point of no use.  
For more details on the build-up effect we refer the reader to~\citep{SB2022_J3}.

\begin{figure}[H]
\centering
\hspace*{-1cm}
\includegraphics[scale=0.22]{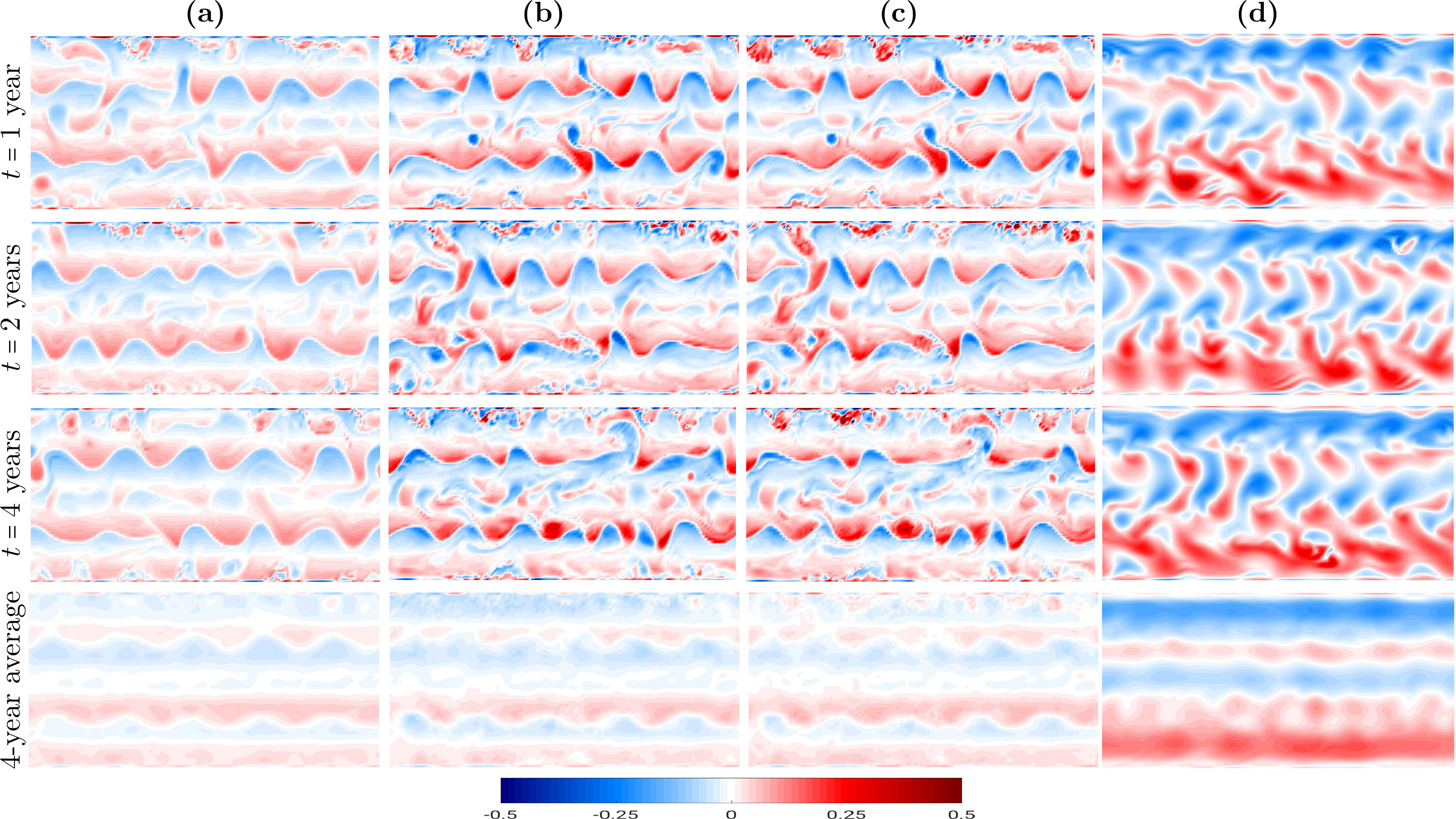}
\caption{Shown are snapshots of the top layer PV anomaly: {\bf (a)} the reference solution (computed on the $513\times257$
grid and then projected onto the $129\times65$ grid), {\bf (b)}/{\bf (c)} probabilistic solution computed with the first/second method, 
{\bf (d)} modelled solution computed with the QG equations~\eqref{eq:pv} on the coarse grid $129\times65$,
and a 4-year time-average (last row); the nudging strength is $\eta=0.1$ for both probabilistic solutions. 
All solutions are given in units of $[s^{-1}f^{-1}_0]$, where $f_0=0.83\times10^{-4}\, {\rm s^{-1}}$ is the Coriolis parameter.
Note that the probabilistic evolutionary methods use only the first 2 years out of the 4-year long reference solution.
Both methods reproduce nominally-resolved on the coarse grid flow features, while the coarse-grid QG solution results in 
complete failure to reproduce even large-scale jets not to mention nominally-resolved features.}
\label{fig:qg_sol}
\end{figure}

In order to avoid the build-up effect we use the nudging methodology. 
We set the nudging parameter as $\eta=0.1$ (it is not the only choice)
in equations~\eqref{eq:ode_y1} and~\eqref{eq:ode_y2}, and present the results in Figure~\ref{fig:qg_sol}.
As seen in Figures~\ref{fig:qg_sol}b,c, both methods reproduce the 
large-scale reference flow structures (two zonally-elongated jets) as well as small-scale vortices and meanders along the jets of the reference solution (Figure~\ref{fig:qg_sol}a).
However, the coarse-grid QG model cannot reproduce the large-scale flow structures not to mention the small-scale ones (Figure~\ref{fig:qg_sol}d).
It is also important to remark that the rapid change of the trajectory computed with the first method (which we observed for the Lorenz system) 
does not seem to reveal itself in the QG dynamics, and both methods give qualitatively the same results. Therefore we further use the second method as 
it is somewhat faster than the first one.

\begin{figure}[H]
\centering
\hspace*{-0.35cm}
\includegraphics[scale=0.22]{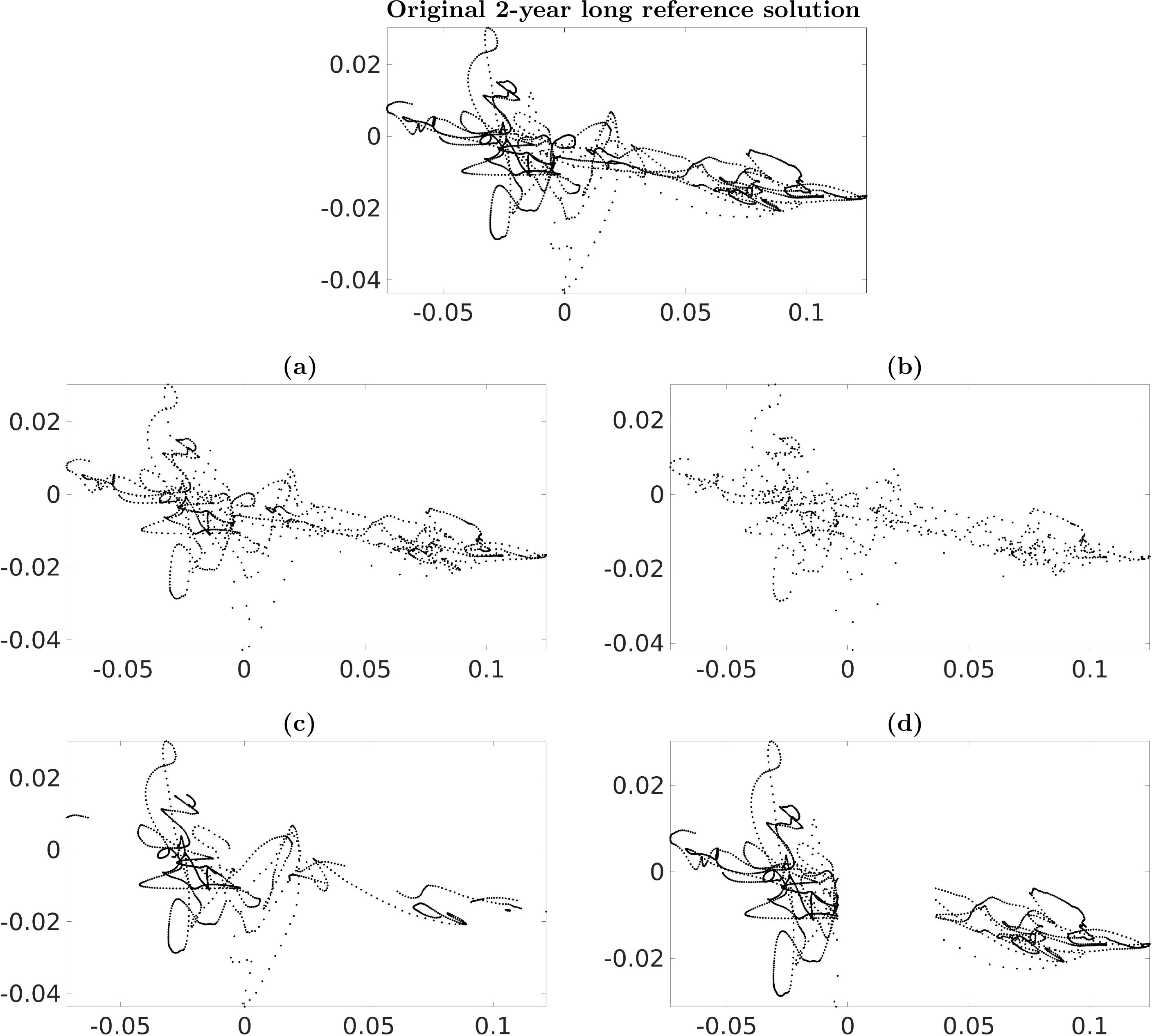}
% \begin{tabular}{cc}
% \multicolumn{2}{c}{\hspace{2cm}\begin{minipage}{0.475\textwidth} {\bf Original 2-year long reference solution} \end{minipage}}\\
% \multicolumn{2}{c}{\hspace*{-3.75cm}\begin{minipage}{0.24\textwidth}\includegraphics[scale=0.1]{q1ref_129x65_DT12_phase_space_coord1_2.jpg}\end{minipage}}\\
% & \\
% \hspace*{1.5cm}\begin{minipage}{0.1\textwidth} {\bf (a)} \end{minipage} & \hspace*{1.5cm}\begin{minipage}{0.1\textwidth} {\bf (b)} \end{minipage}\\
% \hspace*{0cm}\begin{minipage}{0.5\textwidth}\includegraphics[scale=0.1]{{q1ref_129x65_DT12_phase_space_coord1_2__gappy_dynamics_dt2}.jpg}\end{minipage} &
% \hspace*{0cm}\begin{minipage}{0.5\textwidth}\includegraphics[scale=0.1]{{q1ref_129x65_DT12_phase_space_coord1_2__gappy_dynamics_dt4}.jpg}\end{minipage}\\
% & \\
% \hspace*{1.5cm}\begin{minipage}{0.1\textwidth} {\bf (c)} \end{minipage}   & \hspace*{1.5cm}\begin{minipage}{0.1\textwidth} {\bf (d)} \end{minipage}\\
% \hspace*{0cm}\begin{minipage}{0.5\textwidth}\includegraphics[scale=0.1]{{q1ref_129x65_DT12_phase_space_coord1_2__holey_dynamics_r0.979}.jpg}\end{minipage} &
% \hspace*{0cm}\begin{minipage}{0.5\textwidth}\includegraphics[scale=0.1]{{q1ref_129x65_DT12_phase_space_coord1_2__disjoint_dynamics_r0.02}.jpg}\end{minipage}\\
% \end{tabular}
\caption{Shown is a 2D projection of the dynamics in the reference phase space (two coordinates 
from the multi-dimensional reference phase space of the top layer PV anomaly) for the original 2-year long reference solution used in the 
QG simulations above,
{\bf (a)}/{\bf (b)} gappy trajectory with every second/fourth point retained, 
{\bf (c)} holey dynamics (the reference solution contained in a sphere of radius $r=0.979$ centered at its 
time mean is removed), {\bf (d)} disjoint space (the reference phase space is cut into two disjoint regions; the cut width is 0.02).
}
\label{fig:qg_incomplete}
\end{figure}

{\bf Incomplete reference data}. Incomplete observational data are typical in ocean modelling especially when running comprehensive ocean models. 
Obviously, such data cannot be directly used for numerical modelling, as they may include undefined values, missing parts of data records, or combination of both. 
There are different interpolation methods and reanalysis data to overcome the problem. As the probabilistic evolutionary approach is intended to work with comprehensive ocean models, it would be instructive 
to study its performance on incomplete, raw reference flows, i.e. without engaging interpolation or reanalysis. 
For doing so, we take the second method 
% (as it is faster than the first one and gives qualitatively the same results as we have seen above) 
and consider similar to the Lorenz system test cases (Figure~\ref{fig:qg_incomplete}): 
(1)  gappy trajectory, (2) holey dynamics, and (3) disjoint space. We use the same 2-year long reference solution as above,
while running the probabilistic evolutionary model for four years. 

\begin{figure}[H]
\centering
\hspace*{-1cm}
\includegraphics[scale=0.22]{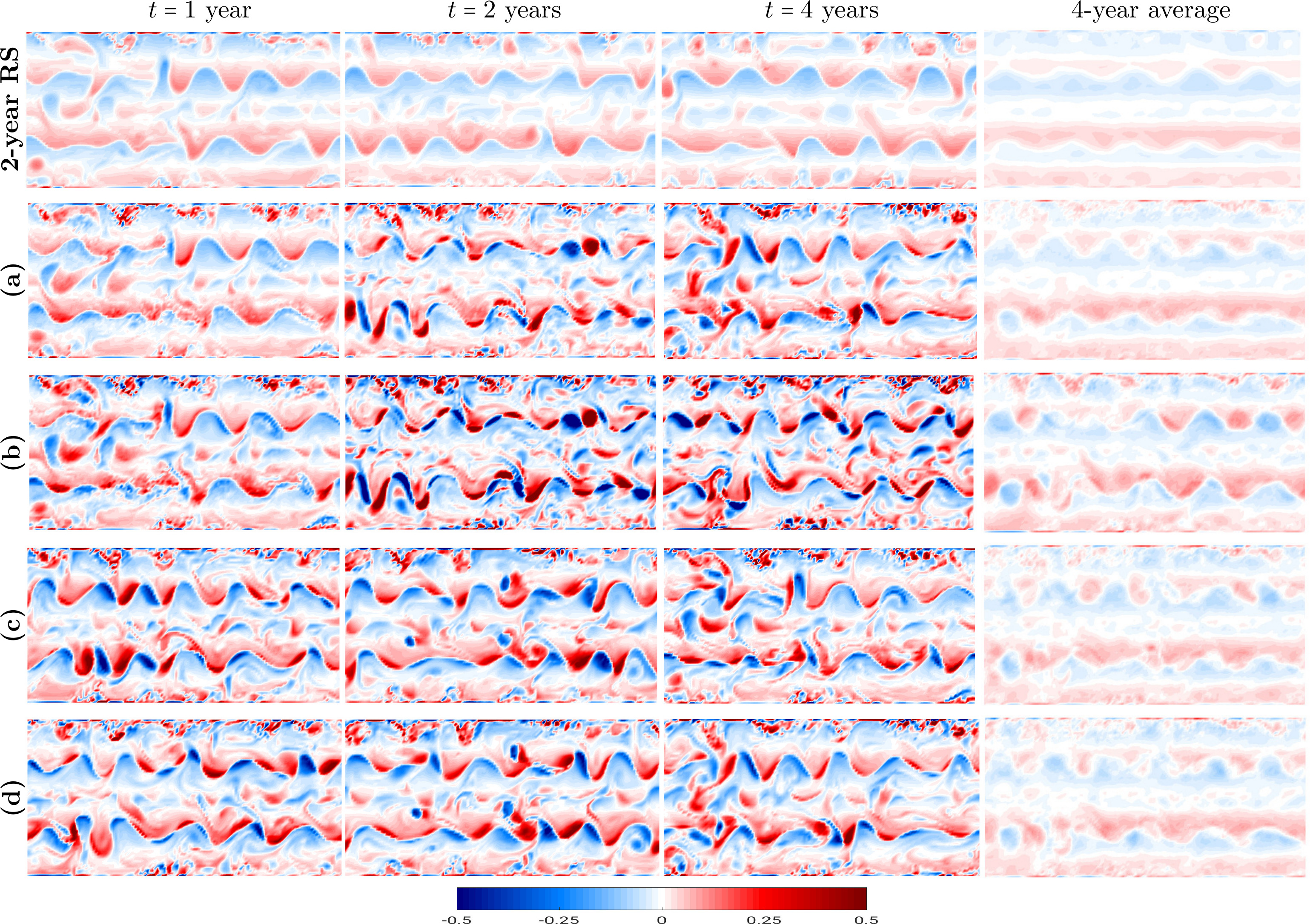}
\caption{Shown are snapshots of the top layer PV anomaly for the incomplete reference solution (Figure~\ref{fig:qg_incomplete}):
{\bf (top row)} the original 2-year long complete reference solution (as in Figure~\ref{fig:qg_sol}a),
{\bf (a)}/{\bf (b)} probabilistic evolutionary dynamics for the gappy reference solution with every second/fourth point removed
(Figure~\ref{fig:qg_incomplete}a,b),
{\bf (c)} probabilistic evolutionary dynamics for the holey reference solution (Figure~\ref{fig:qg_incomplete}c),
{\bf (d)} probabilistic evolutionary dynamics for the disjoint reference solution (Figure~\ref{fig:qg_incomplete}d).
and a 4-year time-average (last column); the nudging strength is $\eta=0.1$. 
All solutions are given in units of $[s^{-1}f^{-1}_0]$, where $f_0=0.83\times10^{-4}\, {\rm s^{-1}}$ is the Coriolis parameter.
Note that the probabilistic evolutionary method reproduces both large- and small-scale flow features even for significantly incomplete reference data sets.
}
\label{fig:qg_incomplete_pv}
\end{figure}

In the {\it gappy trajectory} test case we remove every second (Figure~\ref{fig:qg_incomplete}a) and every fourth (Figure~\ref{fig:qg_incomplete}b)
point from the original 2-year long reference solution thus retaining only 50\% and 25\% of the reference data. 
As seen in Figures~\ref{fig:qg_incomplete_pv}a,b, the probabilistic evolutionary solution reproduces the nominally-resolved reference flow structures
way beyond the time period over which the reference data is available.

The {\it holey dynamics} test case is harder than the previous one as we remove a vast region of the reference dynamics. Namely,
the reference solution contained in the sphere of radius $r=0.979$ centered at its time mean has been excluded from the 
reference solution thus making voids in different parts of the reference trajectory; note that only half of the reference solution remained
after this resection.
As with the previous test case,
the probabilistic evolutionary method restores (Figure~\ref{fig:qg_incomplete_pv}c) the nominally-resolved flow structures 
of the reference solution.

The {\it disjoint space} is the hardest test case in the series, as it cuts the phase space into two disjoint regions (divided by a gap of width 0.02).
Despite that, the probabilistic evolutionary solution is still able to evolve in the reference phase space and reproduce
nominally-resolved reference flow features (Figure~\ref{fig:qg_incomplete_pv}d). 

\begin{figure}[H]
\centering
\hspace*{-1cm}
\includegraphics[scale=0.22]{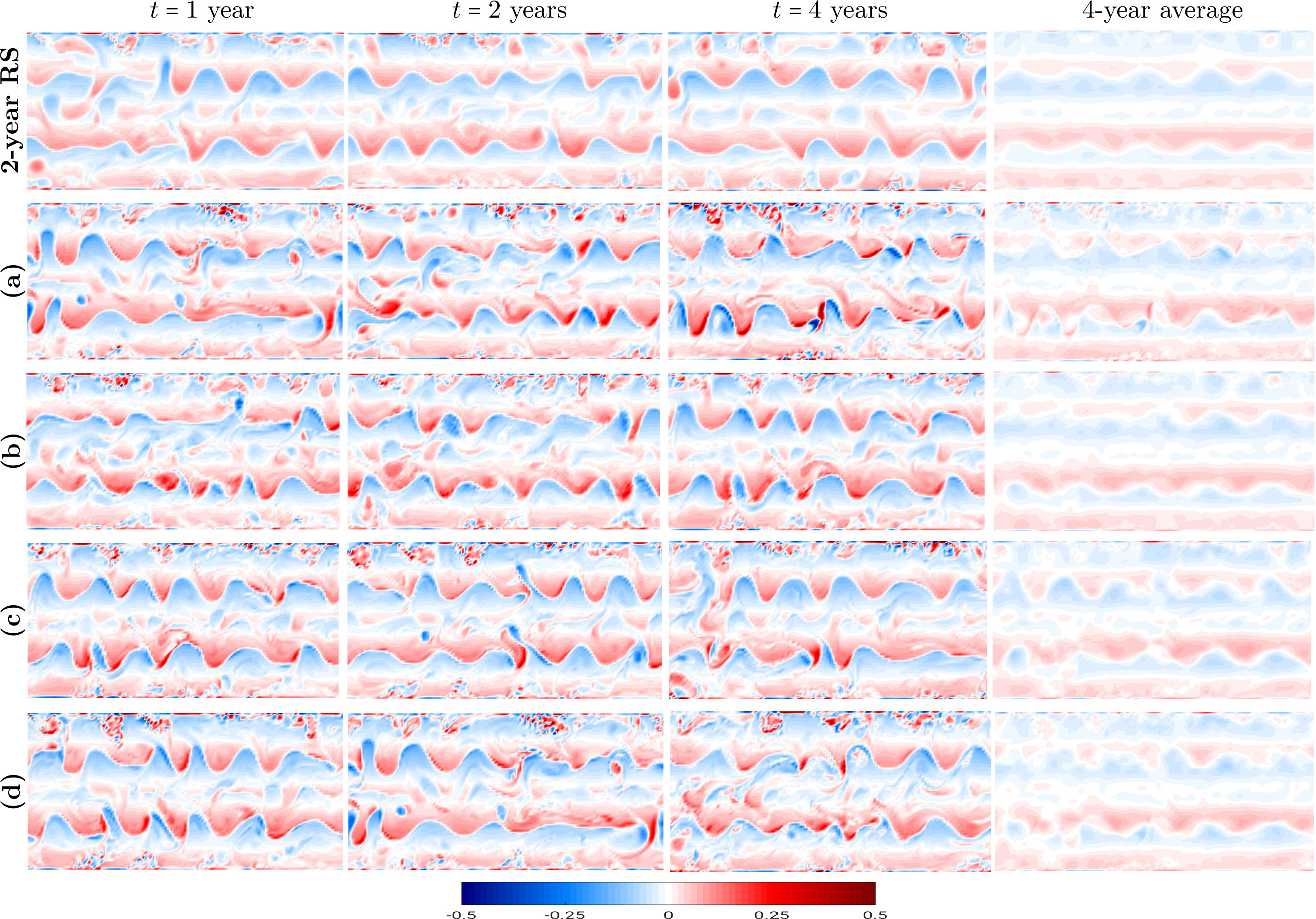}
\caption{The same as Figure~\ref{fig:qg_incomplete_pv} but with the nudging strength $\eta=0.2$ for {\bf (a)}-{\bf (b)}, $\eta=0.6$ for {\bf (c)}, 
and $\eta=0.4$ for {\bf (d)}.
}
\label{fig:qg_incomplete_pv2}
\end{figure}

Although the probabilistic solution reproduces both large- and small-scale reference flow structures for significantly 
corrupted data sets, 
the results in Figure~\ref{fig:qg_incomplete_pv} clearly show that it is overheated, i.e. significantly larger in the absolute value than the reference solution.
It happens because the incomplete reference data works as a repeller thus pressing the probabilistic trajectory out of the reference space.
% In other words, the less data is available for the method to use, the harder it is to keep the trajectory in the reference phase space. 
On the one hand, it might be considered a weakness of the proposed approach, while on the other hand an option to explore vaster regions of the reference phase space,
and thus study probable reference solutions that can potentially be simulated with the reference model. 
In order to keep the probabilistic evolutionary solution in the reference phase space (and thus make its amplitude closer 
to the reference one),
we crank up the nudging strength (Figure~\ref{fig:qg_incomplete_pv2}). It can be done either manually (as in this study) or automatically with 
the adaptive nudging~\citep{SB2022_J1}. The stronger nudging makes the amplitude of all probabilistic solutions 
smaller by keeping them closer to the reference phase space, although we adjusted the nudging strength individually for each case.
Based on the PEA performance for the incomplete reference solutions, we conclude 
that the PEA works well even with significantly corrupted reference data. This is an appealing features not only for ocean modellers working with
models but also for those working with measurements. 

\begin{figure}[H]
\centering
\hspace*{-1cm}
\includegraphics[scale=0.22]{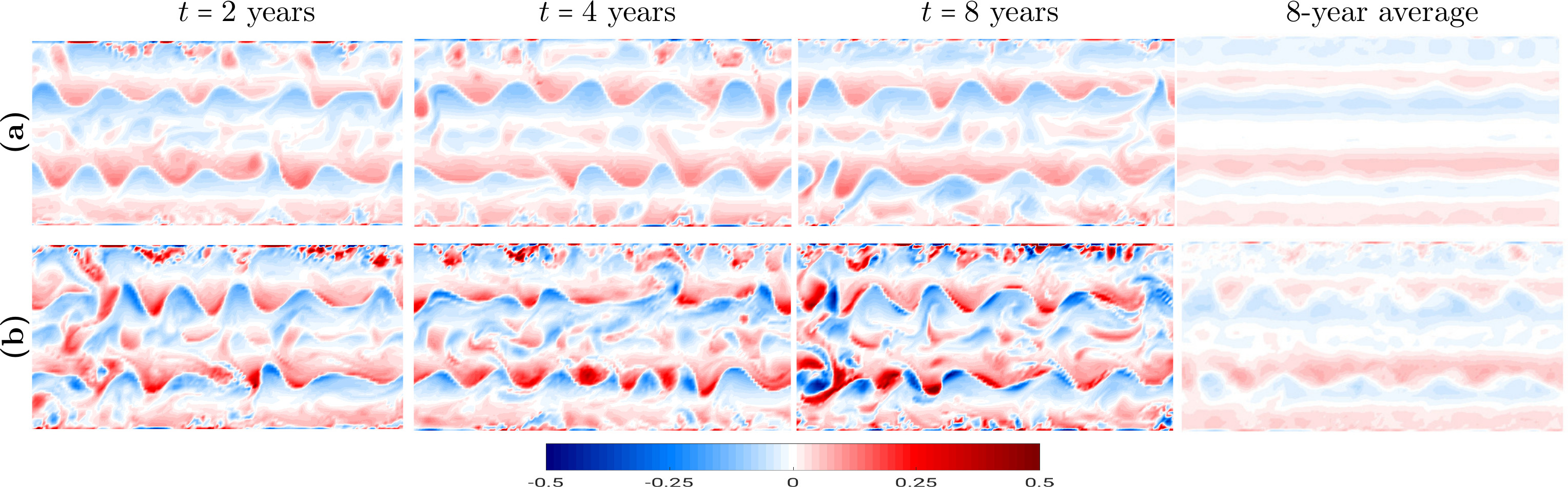}
\caption{
Shown are snapshots of the top layer PV anomaly: {\bf (a)} the reference solution, 
{\bf (b)} probabilistic solution computed with the second method,
and a 8-year time-average (last column); the nudging strength is $\eta=0.1$. 
All solutions are given in units of $[s^{-1}f^{-1}_0]$, where $f_0=0.83\times10^{-4}\, {\rm s^{-1}}$ is the Coriolis parameter.
As with shorter runs, the probabilistic evolutionary method reproduces the nominally-resolved reference flow features (both large and small scales) 
over the period of 8 years.
}
\label{fig:qg_long}
\end{figure}

{\bf Long simulations}. It might seem from the results above that simulations with the PEA can only be twice as long as the reference solution, 
% not because it is enough to demonstrate the ability of the methods to work well beyond the reference data set, 
% but because it is the limit for the proposed approach.
thus setting the upper limit for PEA simulations.
In what follows, we demonstrate how the PEA works on longer time-scales. We take the second method
and run it for 8 years, while using the same 2-year long reference solution (Figure~\ref{fig:qg_long}). As seen in the figure, the method
reproduces nominally-resolved flow features (large-scale jets, small-scale vortices, and meanders along the jets) of the reference solution.
This, once again, ensures that the PEA can model ocean flows far beyond the reference data set.

\begin{figure}[H]
\centering
\hspace*{-1cm}
\includegraphics[scale=0.22]{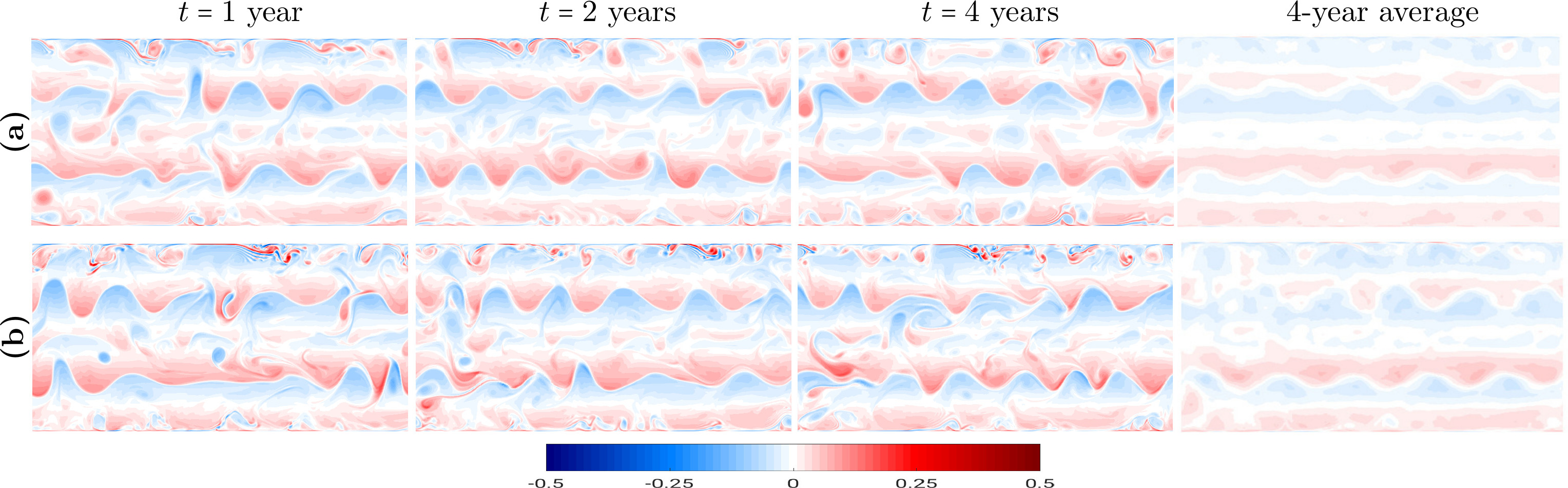}
\caption{Shown are snapshots of the top layer PV anomaly: {\bf (a)} the reference solution (computed on the $513\times257$ grid), 
{\bf (b)} probabilistic solution (computed with the second method on the grid $513\times257$),
and a 4-year time-average (last column); the nudging strength is $\eta=0.1$. 
All solutions are given in units of $[s^{-1}f^{-1}_0]$, where $f_0=0.83\times10^{-4}\, {\rm s^{-1}}$ is the Coriolis parameter.
Note that the probabilistic evolutionary method reproduces the nominally-resolved reference flow features (large-scale jets and small-scale vrotices).}
\label{fig:qg_high_res_sol}
\end{figure}

{\bf The high-resolution simulation}. 
As the PEA is proposed as an alternative to modelling the ocean, it would be instructive to assess it on
high-resolution data as well. In order to do it, we take a 2-year long high-resolution 
QG solution (computed on the grid $513\times257$) as a reference solution, and study how the second method performs 
(Figure~\ref{fig:qg_high_res_sol}). As with the low-resolution reference solution, the method performs equally well and 
reproduces the nominally-resolved reference flow features.

\section{Conclusions and discussion\label{sec:conclusions}}
In this work we have proposed a probabilistic evolution (PEA) approach to ocean modelling
that capitalize on the chaotic nature of ocean dynamics by taking advantage of using
the probability distribution of neighbourhood states in the reference phase space as opposed to making use of 
deterministic or stochastic differential equations.
A new state of the model is determined by the likelihood of the states neighbouring to the current state. 
The probabilistic nature of the flow evolution implies that even very unlikely (rare) events are expected to occur once in a while
thus echoing our observations of extreme weather and climate events.

% tests
Within the PEA framework we have developed two probabilistic evolutionary methods. These methods have been tested
on the Lorenz 63 system and showed that both methods reproduce the Lorenz attractor even for significantly corrupted reference data sets.
In addition, we have showed how the nudging strength can influence the probabilistic solution. Being assured in the PEA potential for modelling geophysical flows,
we have considered an idealized ocean model (two-layer quasi-geostrophic model configured for a horizontally periodic flat-bottom channel)
and showed that a non-eddy-resolving solution can be significantly improved towards the reference eddy-resolving solution. Within the context of QG dynamics
we have demonstrated the build-up effect and its detrimental consequences on the probabilistic solution, and how to avoid it with using the nudging methodology.
We have also studied how the probabilistic evolutionary models work on incomplete reference data and demonstrated that they reproduce nominally-resolved on the coarse 
grid reference flow features even for significantly corrupted reference data sets. In addition, we have demonstrated that the PEA works very well over long time periods
(8 years in our case) even for short reference solutions (2-year long). 
Our results show that the probabilistic evolutionary approach performs equally well for both 
low- and high-resolution reference solutions. 

% benefits
The appealing advantages of the probabilistic evolutionary approach are: (1) it requires no modification of the ocean model;
(2) easy to implement; (3) it can take not only the reference solution as input data
but also real measurements from different sources (drifters, weather stations, etc.), or combination of both;
(4) it is ready out of the box for generating ensembles of solutions, (5) copes with significantly corrupted data sets,
(6) performs equally well for both low- and high-resolution reference data, and reproduces nominally-resolved reference dynamics;
(7) works over long time-scales without degradation of the 
probabilistic solution (even for short reference records) thus operating well beyond the reference data range.

All this offers a great flexibility to ocean modellers working with comprehensive ocean models and measurements, and 
allow us to expect that 
the proposed approach has strong potential for the use in the context of primitive equations which we plan to approach in the future research.

% future research
% Some important questions are left for future research. Namely, how to reconstruct a probabilistic evolutionary model based on the reference data
% and whether the injection of the new data (computed with the PEA) into the pool of available reference data
% will improve the accuracy of probabilistic evolutionary models.

\section{Acknowledgments}
The authors thank The Leverhulme Trust for the support of this work through the grant RPG-2019-024.



%% References


% \bibliographystyle{elsarticle-num}
\bibliographystyle{apalike}
\bibliography{refs}

\begin{thebibliography}{}

\bibitem[Chassignet et~al., 2007]{hycom2007}
Chassignet, E., Hurlburt, H., Smedstad, O., Halliwell, G., Hogan, P.,
  Wallcraft, A., Baraille, R., and Bleck, R. (2007).
\newblock The hycom (hybrid coordinate ocean model) data assimilative system.
\newblock {\em Journal of Marine Systems}, 65:60--83.

\bibitem[Cotter et~al., 2020]{CCHPS2020_J2}
Cotter, C., Crisan, D., Holm, D., Pan, W., and Shevchenko, I. (2020).
\newblock Modelling uncertainty using stochastic transport noise in a 2-layer
  quasi-geostrophic model.
\newblock {\em Foundations of Data Science}, 2:173--205.

\bibitem[Danilov et~al., 2017]{Danilov_etal_2017}
Danilov, S., Sidorenko, D., Wang, Q., and Jung, T. (2017).
\newblock The {F}inite-volum{E} {S}ea ice–{O}cean model (fesom2).
\newblock {\em Geosci. Model Dev.}, 10:765--789.

\bibitem[Devroye, 1986]{Devroye1986}
Devroye, L. (1986).
\newblock {\em Non-Uniform Random Variate Generation}.
\newblock Springer-Verlag, New York, Berlin, Heidelberg, Tokyo.

\bibitem[Lorenz, 1963]{Lorenz1963}
Lorenz, E. (1963).
\newblock Deterministic nonperiodic flow.
\newblock {\em J. Atmos. Sci.}, 20:130--141.

\bibitem[Madec and {NEMO System Team}, 2022]{NEMO2022}
Madec, G. and {NEMO System Team} (2022).
\newblock Nemo ocean engine.
\newblock {\em Scientific Notes of Climate Modelling Center, Institut
  Pierre-Simon Laplace}, 27.

\bibitem[Marshall et~al., 1997]{Marshall_etal_1997}
Marshall, J., Adcroft, A., Hill, C., Perelman, L., and Heisey, C. (1997).
\newblock A finite-volume, incompressible {N}avier {S}tokes model for studies
  of the ocean on parallel computers.
\newblock {\em J. Geophys. Res.}, 102:5753--5766.

\bibitem[McWilliams, 1977]{McWilliams1977}
McWilliams, J. (1977).
\newblock A note on a consistent quasigeostrophic model in a multiply connected
  domain.
\newblock {\em Dynam. Atmos. Ocean}, 5:427--441.

\bibitem[Pedlosky, 1987]{Pedlosky1987}
Pedlosky, J. (1987).
\newblock {\em Geophysical fluid dynamics}.
\newblock Springer-Verlag, New York.

\bibitem[Shevchenko and Berloff, 2021]{SB2021_J1}
Shevchenko, I. and Berloff, P. (2021).
\newblock A method for preserving large-scale flow patterns in low-resolution
  ocean simulations.
\newblock {\em Ocean Model.}, 161:101795.

\bibitem[Shevchenko and Berloff, 2022a]{SB2022_J3}
Shevchenko, I. and Berloff, P. (2022a).
\newblock A hyper-parameterization method for comprehensive ocean models:
  Advection of the image point.
\newblock {\em arXiv:2209.07141}.

\bibitem[Shevchenko and Berloff, 2022b]{SB2022_J2}
Shevchenko, I. and Berloff, P. (2022b).
\newblock A method for preserving nominally-resolved flow patterns in
  low-resolution ocean simulations: Constrained dynamics.
\newblock {\em Ocean Model.}, 178:102098.

\bibitem[Shevchenko and Berloff, 2022c]{SB2022_J1}
Shevchenko, I. and Berloff, P. (2022c).
\newblock A method for preserving nominally-resolved flow patterns in
  low-resolution ocean simulations: Dynamical system reconstruction.
\newblock {\em Ocean Model.}, 170:101939.

\end{thebibliography}



\end{document}